\documentclass[conference]{IEEEtran}
\IEEEoverridecommandlockouts
\usepackage{cite}
\usepackage{amsmath,amssymb,amsfonts}
\usepackage{graphicx}
\usepackage{textcomp}
\usepackage{xcolor}
\usepackage{enumitem}
\usepackage{lipsum}
\usepackage{tabularx}
\usepackage{algorithm}
\usepackage{algpseudocode}

\def\BibTeX{{\rm B\kern-.05em{\sc i\kern-.025em b}\kern-.08em
    T\kern-.1667em\lower.7ex\hbox{E}\kern-.125emX}}
\begin{document}

\title{TruChain: A Multi-Layer Architecture for Trusted, Verifiable, and Immutable Open Banking Data}

\author{
\IEEEauthorblockN{
    Aufa Nasywa Rahman, Bimo Sunarfri Hantono, and Guntur Dharma Putra\IEEEauthorrefmark{1}
}
\IEEEauthorblockA{
    Universitas Gadjah Mada, Indonesia \\
    aufa.nas2003@mail.ugm.ac.id, \{bhe, gdputra\}@ugm.ac.id
}
\thanks{\IEEEauthorrefmark{1}Corresponding author.}
}

\maketitle

\begin{abstract}
Open banking framework enables third party providers to access financial data across banking institutions, leading to unprecedented innovations in the financial sector. However, some open banking standards remain susceptible to severe technological risks, including unverified data sources, inconsistent data integrity, and lack of immutability. In this paper, we propose a layered architecture that provides assurance in data trustworthiness with three distinct levels of trust, covering source validation, data-level authentication, and tamper-proof storage. The first layer guarantees the source legitimacy using decentralized identity and verifiable presentation, while the second layer verifies data authenticity and consistency using cryptographic signing. Lastly, the third layer guarantees data immutability through the Tangle, a directed acyclic graph distributed ledger. We implemented a proof-of-concept implementation of our solution to evaluate its performance, where the results demonstrate that the system scales linearly with a stable throughput, exhibits a 100\% validation rate, and utilizes under 35\% of CPU and 350 MiB memory.
Compared to a real-world open banking implementation, our solution offers significantly reduced latency and stronger data integrity assurance. Overall, our solution offers a practical and efficient system for secure data sharing in financial ecosystems while maintaining regulatory compliance. 
\end{abstract}

\begin{IEEEkeywords}
open banking, data integrity, verifiable credentials, tangle, trust architecture, secure data sharing
\end{IEEEkeywords}

\section{Introduction}
The rise of immersive digital environments, like the metaverse, is transforming how individuals interact, own digital assets, and perform economic transactions. These transformations allow for novel activities such as purchasing virtual goods and securing virtual property, which all require secure and verifiable financial transactions that bridge real-world banking and digital ecosystems~\cite{2024IqbalMetaverse}. Assurance in the authenticity and integrity of financial data in such transactions is essential for ensuring trust, credit, identity, and risk assessments~\cite{pwc2022metaverse}. In addition, the financial industry is undergoing rapid digital transformations, driven by mobile money, digital banks, and financial technology services, which leads to improved system connectivity and data usability, initiating the growth of open banking~\cite{2023Fintech-and-The-Future-Of-Finance, 2023Fintech-and-Digital-Transfromation}.

Open banking is a regulatory and technological framework that enables secure exchange of consumer-permissioned financial data across institutions and third-party providers, which is commonly done via standardized APIs~\cite{2019Digital-Frontiers-Institute}. The European Union’s Payment Services Directive 2 (PSD2) was a pioneering regulatory instrument, while subsequent global developments have introduced various governance models, including regulation-led frameworks in the UK and Brazil, which mandate data-sharing obligations and technical standards~\cite{2019Digital-Frontiers-Institute}. Another market-driven approach is coined in the US, where participation is voluntary~\cite{2019Digital-Frontiers-Institute}.
PSD2, in particular, provides a structured precedent for secure, interoperable, and consumer-centric systems, which introduces two main roles, namely Account Information Service Provider (AISP) and Payment Initiation Service Provider (PISP). In general, AISP provides services that aggregate financial account information from multiple banks with consumer consent, while PISP facilitates online payments by initiating transfers directly from a payer’s bank account to a payee’s account~\cite{2024PSD-AlgaraMunoz2024}. However, in increasingly complex digital financial ecosystems, some challenges remain unsolved, particularly in verifying the data sources and ensuring data authenticity, which are instrumental in providing trusted, verifiable, and immutable financial services. In the subsequent paragraphs, we discuss these challenges in more detail.

\textit{Data source verification.} Almehrej et al.~\cite{2020Almehrej-PSD2Security} revealed that PSD2 framework relies heavily on the assumption that data sources are fully trusted. As such, in the absence of reliable data source verification, the data that being ingested on the PSD2 framework, specifically within the AISP role, may originate from malicious or unreliable sources which could compromise the entire data integrity due to questionable data accuracy and fraudulent, or intentional manipulations.

\textit{Data authenticity and consistency.} Even though PSD2 already implements secure protocol like TLS, it does not guarantee data-level verification~\cite{2024HarmonizingPSD2}. On the other hand, the issue of data inconsistencies may arise due to technical differences and minimum harmonization between online platforms or data sources. These conditions could significantly compromise data integrity, even if the source of the data is authenticated. For instance, financial data could be inconsistent, tampered or corrupted during the transmission process between authenticated parties.

\textit{Data immutability.} Over time, data may be subject to tampering or unauthorized modifications. In conventional centralized systems, malicious actors, whether internal or external, can alter historical financial records without immediate detection, compromising the integrity of the data. For example, a compromised financial institutions could manipulate a user's transaction history to suppress evidence of recurring overdrafts or inflate account balances before forwarding the data. Such manipulation could falsely enhance the user’s creditworthiness, leading to fraudulent loan approvals or misinformed credit decisions~\cite{2021fabcic-sca}. PSD2 framework lacks a verifiable data integrity mechanism, leaving it vulnerable to such historical data tampering~\cite{2021fabcic-sca}.

In this paper, we propose a layered architecture to address data trustworthiness issues in open banking. 
Our architecture consists of three interconnected layers, where each layer is responsible for addressing a specific challenge. The first layer verifies the data source before proceeding with data transmission, while the second layer provides assurance in data consistency during transmission. Lastly, the third layer guarantees data immutability at the final stage to provide tamper-proof records that establish trust among all participating entities in the open banking ecosystem. In practice, our solution begins with the data source sending its credentials to an issuer for validation. Upon successful validation, the data source receives Verifiable Credentials (VCs) anchored to the issuer’s and source’s Decentralized Identities (DIDs). When the data source needs to send data, it presents the VC as a Verifiable Presentation (VP) to the first layer, along with a corresponding challenge. Once validated, the source is authorized to send data for a limited amount of time. In the second layer, the data undergoes processing to verify its integrity and consistency. If the data passes validation, it proceeds to the third layer, where the data is made immutable by storing its hash on-chain while the raw data is stored off-chain with its corresponding block ID.

In summary, the contributions of this paper are three-fold:
\begin{itemize}
    \item We propose a three-layered architecture to address trust issues in the data exchange mechanism under the open banking framework, where each layer is designed to tackle a specific issue.

    \item We develop a source authentication mechanism integrated with data ingestion process to verify data source. We incorporate a decentralized authority with decentralized DID infrastructure to significantly reduce attack possibilities. 

    \item We introduce a mechanism to ensure data authenticity, integrity and immutability implemented by using cryptographic signing technique within decentralized infrastructure.
\end{itemize}

The rest of the paper is organized as follows, We motivate our solution in Section~\ref{sec:motivation} and review related work in Section~\ref{sec:related-work}. We outline our solution in Section ~\ref{sec:system-model}, while Section~\ref{sec:implementation-and-performance} discusses the performance evaluation. We conclude our paper in Section~\ref{sec:conclusion}.

\section{Motivation}
\label{sec:motivation}

\begin{figure}[t]
    \centering
    \includegraphics[width=0.9\linewidth]{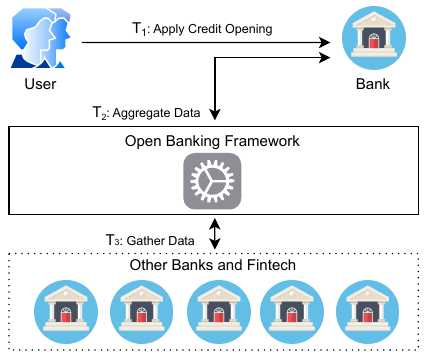}
    \caption{Example of a credit scoring application, where the user acts as the credit applicant and the bank aggregates financial data from multiple banks and financial technology providers using the open banking framework.}
    \label{fig:use-case}
\end{figure}

Open banking is increasingly being adopted to support alternative approaches to credit risk assessment, especially for individuals with limited credit histories or non-traditional financial profiles. As noted by Zetzsche et al.~\cite{zetzsche2019future}, the evolution of regulatory and technological infrastructures has accelerated a shift away from static, legacy-based credit scoring models toward more granular, real-time assessments enabled by financial data portability and accessibility. 

Fig.~\ref{fig:use-case} depicts a credit scoring example involving multiple entities: 1) user; 2) bank; 3) open banking framework; and 4) other banks and financial technology. The process begins with the user initiating a credit application ($T_1$: Apply Credit Opening) at their primary bank (user bank). In response, the user bank accesses the applicant's financial data from other banks and financial technology providers ($T_2$: Aggregate Data). The data aggregated was sourced from multiple banks and financial institutions ($T_3$: Gather Data), such inter-institutional access is made possible through the underlying open banking framework, which facilitates standardized API-based data sharing with user consent. Data retrieved may include account balances, income streams, spending patterns, and transaction histories which are critical inputs for assessing creditworthiness in data-driven financial services.

However, while the open banking framework enables access, it does not inherently guarantee that the collected data is authentic, originates from a legitimate source, or remains unaltered in transit. In credit scoring workflows, such trust assurances are crucial, as any tampering, omission, or falsification of financial data can lead to inaccurate risk assessments and undermine the fairness and reliability of lending decisions. To address these concerns, a trusted data layer is required to ensure that each financial record can be traced to its verified source, cryptographically validated, and logged in a tamper-evident manner. 

\section{Related Work}
\label{sec:related-work}
Waliullah et al.~\cite{waliullah2025cybersecurity} conducted a systematic literature review that analyzed cybersecurity threats affecting digital banking, with specific attention to regulatory frameworks such as PSD2 which identifies persistent threats such as phishing, malware, and data breaches as major obstacles to digital banking adoption and trust. While PSD2 mandates strong customer authentication (SCA) and API-based interoperability, the review highlights that compliance alone is insufficient to address evolving threats. Importantly, the integration of third-party providers introduces additional security risks, requiring more advanced, layered cybersecurity measures. 

The paper by Polasik et al.~\cite{polasik2024evaluating} offers an extensive empirical assessment of PSD2, positioning it as both a driver of innovation and a source of unresolved regulatory and technical challenges in the European open banking landscape. Polasik et al.~\cite{polasik2024evaluating} acknowledge that PSD2 has been instrumental in enabling new market entrants and creating an open API-based infrastructure. However, they emphasize that the associated implementation, especially for SCA and API integration have far outweighed the tangible benefits so far. Additionally, the absence of a harmonized API standard and uneven national licensing practices have led to market fragmentation and regulatory arbitrage, particularly concentrated in jurisdictions like Lithuania. These shortcomings undermine the directive’s original goals of transparency, consumer empowerment, and competition. Polasik et al.~\cite{polasik2024evaluating} suggesting clearer boundaries between payment and data access services, improved supervision, and further infrastructure standardization. 

Recent advancements in decentralized identity and trust architecture have led to several approaches for securing data exchange across distributed systems. Kovach et al.~\cite{2024Kovach-SSI} proposed a comprehensive architecture for sovereign data exchange in Industrial Internet of Thing (IIoT). The architecture built by integrating IOTA Identity, consisting of IOTA's Tangle-based Distributed Ledger Technology (DLT), VC, and the International Data Spaces (IDS) reference architecture. Their works demonstrates strong capabilities in enforcing identity, usage control, and data integrity in industrial environments. Trust is established through the use of DIDs for authenticating nodes, VCs for attaching signed metadata, and immutable logging on the IOTA Tangle.

Ghosh et al.~\cite{ghosh2023consenTrack} proposes a blockchain-based solution designed to address the challenges of consent management in open banking. Despite the potential of open banking, customer adoption has been limited due to concerns regarding data privacy and transparency. Framework introduced by Ghosh et al~\cite{ghosh2023consenTrack} utilizes blockchain technology to provide a transparent and immutable record of customer consent, thus allowing real-time monitoring of data sharing activities. The blockchain's inherent properties of transparency and immutability support the verification of consent data and the detection of any violations. While the framework does not directly manage the data-sharing mechanism between banks and TPPs, it enhances the transparency and accountability of the consent management process.

Claudio et al.~\cite{2023Claudio-DID-IOTA} introduce a novel IOTA-based DID method integrated into OpenSSL, thus making decentralized identity handling being processed directly on the Tangle. While these systems demonstrate the potential of decentralized infrastructure, they are focused on industrial infrastructure ecosystems and tightly coupled with domain-specific framework like IDS. Moreover, their trust models are anchored at the identity layer alone and lack of a comprehensive, end-to-end multi-layered authentication strategy.

Collectively, the reviewed literature reveals several critical gaps in the current open banking ecosystem. Waliullah et al.~\cite{waliullah2025cybersecurity} highlight persistent cybersecurity threats that existing regulations like PSD2 only partially address, particularly with the integration of third-party providers. Polasik et al.~\cite{polasik2024evaluating} underscore that while PSD2 initiating a huge innovation, its implementation, fragmented API standards, and uneven supervisory practices undermine its effectiveness. Meanwhile, blockchain-based frameworks proposed by Ghosh et al.~\cite{ghosh2023consenTrack} improve consent transparency but do not address data authenticity or source validation. Decentralized identity systems proposed by Kovach et al.~\cite{2024Kovach-SSI} and Claudio et al.~\cite{2023Claudio-DID-IOTA} focus on industrial domains or identity-level trust alone, lacking a holistic data pipeline perspective.

To address these shortcomings, our work introduces a trust layer designed to enforce trust at multiple critical stages in the data life cycle: (1) verifying the legitimacy of data sources, (2) ensuring the authenticity and consistency of transmitted data via cryptographic signatures, and (3) guaranteeing data immutability through anchoring in a DLT.

\begin{figure*}[t]
    \centering
    \includegraphics[width=0.95\linewidth]{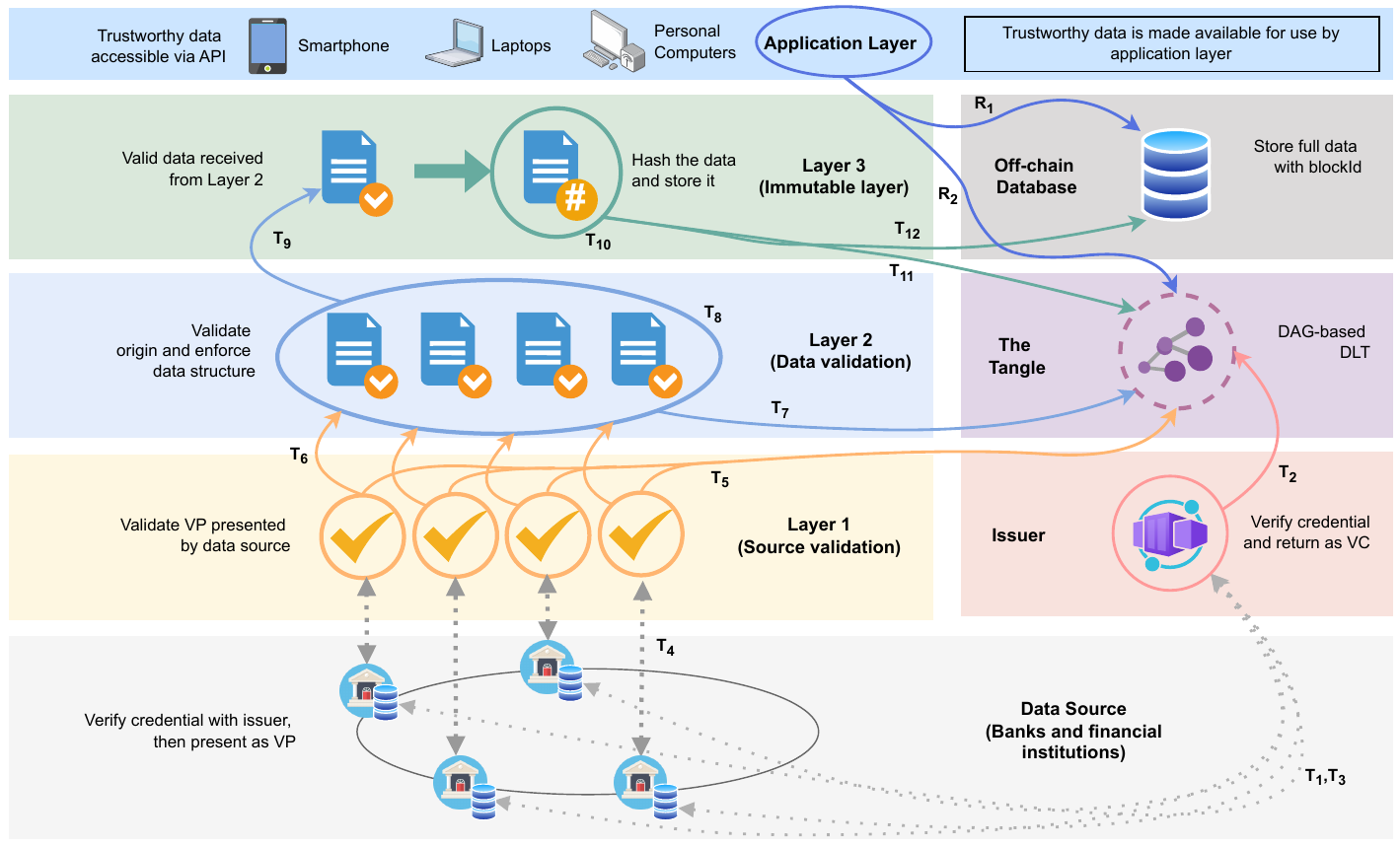}
    \caption{Proposed three-layer trust architecture, showing interactions in two cases: storing data (T) and querying data (R). Storing data begins with the data source validating its credentials with the issuer and proving its trust to the trust layer. Once verified, it is allowed to submit data for further processing. In the query process, data is first retrieved from the off-chain database, then its hash is compared with the corresponding on-chain hash. If they match, the data is considered valid.}
    \label{fig:proposed-layer-detailed}
\end{figure*}

\section{System Model}
\label{sec:system-model}

In this section, we present the conceptual design and core algorithms of the proposed trust layer. We describe how the system enforces trust through a structured sequence of trust validation steps. 

\subsection{System Concept}
\label{subsub:system-concept}

We propose a three-layer trust architecture as illustrated in Fig.~\ref{fig:proposed-layer-detailed}, where each layer performs a specific and distinct function.
The first layer focuses on capturing and validating the source by using source credentials bound to a verifiable presentation.
The second layer handles the validation of data, ensuring that the data is transmitted in verifiable credential format from trusted sources. The second layer also verifies that the data structure adheres to the  pre-defined standard. The third layer is responsible for the immutable storage of verified data by appending its hash to a tamper-proof distributed ledger. The full data is stored off-chain in a NoSQL database that offers scalability and high availability, while only the cryptographic hash is recorded on-chain. Off-chain storage is adopted to ensure compliance with open banking and GDPR requirements, as it allows the use of conventional databases that enforce retention policies. Regulations such as GDPR mandate organizations to handle data subject requests, like access logs and data portability.

There are multiple actors involved in the proposed trust layer, described as follows
\begin{itemize}
    \item \textbf{Trust Layer:} Comprising three layers that ensure trust in the data being ingested.
    
    \item \textbf{Data Source:} Data sources such as banks and financial institutions providing the data.
    
    \item \textbf{Issuer:} Authority responsible for validating the identity of the data sources and generating VC for them.
    
    \item \textbf{The Tangle:} DLT architecture that serves as the trusted source for the system.
    
    \item \textbf{Off-chain Database:} A database used to store full data along with the associated block ID from the blockchain.
    
    \item \textbf{Application Layer:} Interface between the user and the system, provided through various platforms.
\end{itemize}

Meanwhile, the interactions between actors can be organized into several groups, with each group consisting a set of detailed interactions. These interactions can be described as follows
\begin{enumerate}
    \item \textbf{Identity issuance:} Establishes the foundational trust relationship between a data source and the trust architecture, with the issuer serving as the identity verifier. Three primary actors are involved in this process are 1) Data Source; 2) Issuer; and 3) The Tangle. The process consists of the following detailed interactions
    \begin{itemize}
        \item $T_1$: The data source submits its identity information, such as legal documents and its DID, to the issuer for verification.
        
        \item $T_2$: Upon successful verification, the issuer issues a VC, cryptographically linking both parties DIDs and anchoring it to the the Tangle.
        
        \item $T_3$: The issuer returns the VC to the data source, which can now use it as proof of identity in future interactions within the trust layer.
    \end{itemize}

    \item \textbf{Source verification:} The interaction shifts from the issuer to the trust architecture, specifically Layer 1, which is responsible for verifying the authenticity of the data source. Three key actors are involved in this process are 1) Data Source; 2) Layer 1 on trust layer; and 3) The Tangle. The process consists of the following detailed interactions
    \begin{itemize}
        \item $T_4$: The data source requests a cryptographic challenge from Layer 1 to begin verification. Layer 1 responds with a 16-byte random nonce. The data source creates a VP by combining its VC with the challenge, signs it, encodes it as a JWT, and sends it to Layer 1 for validation.
    
        \item $T_5$: Layer 1 validates the VP by checking the digital signature and ensuring the DID matches the one anchored on the the Tangle. If successful, Layer 1 issues a time-limited session token to the data source, enabling it to proceed. Expired tokens require re-verification.
    \end{itemize}
    
    \item \textbf{Data verification:} In this stage, the data source submits verifiable data into the trust architecture for validation. The process involves checking both the authenticity and structure of the submitted information before it can be stored. The key actors are 1) Data Source; 2) Layer 1 on trust layer; and 3) Layer 2 on trust layer. The process consists of the following detailed interactions
    \begin{itemize}
        \item $T_6$: The data source issue a new VC, embedding the actual data, then submits this VC encoded as a JWT alongside with the session token. The VC pass through Layer 1 session token validation then passed to Layer 2.
    
        \item $T_{7}$: Layer 2 decode the VC, verifies its integrity with the DID anchored to the Tangle. 

        \item $T_{8}$: If valid, Layer 2 validate the data structure, and forwards it to Layer 3.
    \end{itemize}

    \item \textbf{Storing verified data:} After passing validation, the verified data is securely stored using both on-chain and off-chain mechanisms. The key actors involved are: 1) Layer 3 on trust layer; 2) The Tangle; and 3) Off-chain Storage. The process consists of the following detailed interactions
    \begin{itemize}
        \item $T_{9}$: Layer 3 receive the trusted data from Layer 2.

        \item $T_{10}$: Trusted data is hashed.
        
        \item $T_{11}$: The hash is submitted to the Tangle. Upon success, a block ID is returned for reference.
    
        \item $T_{12}$: The full data is stored in off-chain database alongside the block ID returned from the Tangle.
    \end{itemize}

    \item \textbf{Query verified data:} In this stage, the trust layer allows data retrieval by checking both on-chain (the Tangle) and off-chain storage. The data is validated by comparing hashes. The key actors involved are: 1) Layer 3 on trust layer; 2) Off-Chain Database; and 3) The Tangle. The process consists of the following detailed interactions
    \begin{itemize}
        \item $R_{1}$: The query first checks off-chain storage for the data. If found, the block ID is retrieved along with the data.
    
        \item $R_{2}$: The block ID is then used to fetch the hash from the Tangle. Then, the full data retrieved from off-chain storage is hashed (excluding the block ID). The computed hash is compared with the hash from the Tangle. If they match, the data is valid.
    \end{itemize}
\end{enumerate}

\subsection{System Protocol and Algorithm}
\label{sub:system-protocol-algorithm}

\subsubsection{\textbf{DID Generation}} 
The complete process of DIDs generation is described in Algorithm~\ref{alg:did_gen}. The symbol $s$ denotes a randomly generated 256-bit scalar used as the private key for cryptographic operations, while $A$ represents the public key derived by scalar multiplication of $s$ with the elliptic curve base point $B$, expressed as $A = [s]B$. The variable $\text{hash}$ refers to the Blake2b-256 hash output applied to the public key $A$, which is then encoded into a human-readable address, $\text{Addr}_{\text{Bech32}}$, using the Bech32 format that includes a human-readable part (HRP). The DID Document, denoted as $D$, is a structured object containing the DID identifier $\texttt{id}$, formatted as \texttt{did:iota:}\(\langle\text{networkHrp}\rangle:\langle\text{uniqueId}\rangle\), along with the verification and assertion methods. The alias output $O_{\text{alias}}$ represents the DID Document on the Tangle, and $TX_{\text{publish}}$ is the signed transaction containing this alias output, ready for submission to the network. Finally, $\sigma$ is the EdDSA (Ed25519) digital signature created by signing a message $M$ with the private key $s$, providing cryptographic proof of identity.

\begin{algorithm}
    \caption{DID Generation and Anchoring}\label{alg:did_gen}
    \begin{algorithmic}[1]
        \State \textbf{Key Generation:}
            \State $s \gets \text{random } 256\text{-bit scalar}$ \Comment{Private key}
            \State $A \gets [s]B$ \Comment{Public key derivation} \label{eq:pubkey}
            
        \State \textbf{Address Creation:}
            \State $\text{hash} \gets \text{Blake2b-256}(A)$
            \State $\text{Addr}_{\text{Bech32}} \gets \text{Bech32Encode}(\text{HRP}, \text{hash})$
            
        \State \textbf{DID Document Construction:}
            \State $D \gets \{\text{id}, \text{verificationMethod}, \text{assertionMethod}\}$
            \State $\text{id} \gets \texttt{did:iota:}\langle \text{networkHrp} \rangle:\langle \text{uniqueId} \rangle$
            
        \State \textbf{Tangle Anchoring:}
            \State $O_{\text{alias}} \gets \text{CreateAliasOutput}(D, \text{Addr}_{\text{Bech32}})$
            \State $TX_{\text{publish}} \gets \text{Sign}(\text{secretManager}, O_{\text{alias}})$
            \State \text{Submit } $TX_{\text{publish}}$ \text{ to IOTA network}
            
        \State \textbf{Assertion Generation:} \Comment{For signing messages}
            \State $\sigma \gets \text{EdDSA.Sign}(s, M)$ \Comment{Using Ed25519}
    \end{algorithmic}
\end{algorithm}

\subsubsection{\textbf{VC Generation and Validation}}
The process of generating and validating VC is described in Algorithm~\ref{alg:vc_generation}. $DID_{\text{issuer}}$ represents the decentralized identifier of the credential issuer; $\text{credential\_data}$ refers to the claims about the credential subject. $VC$ denotes the verifiable credential object containing fields such as a unique identifier, type, issuer, and credential subject. $Header$ is the JWT header specifying the signature algorithm (\texttt{EdDSA}), token type (\texttt{JWT}), and key identifier ($\text{kid}$). $Payload$ is the Base64URL-encoded $VC$, while $signingInput$ is the concatenation of the encoded header and payload separated by a period. $\sigma$ represents the Ed25519 signature generated over the signing input using the issuer's private key $s_{\text{issuer}}$. $J\_signedVC$ is the final signed JWT combining the signing input and the encoded signature. In validation, $Header$, $Payload$, and $\sigma$ are extracted from $J\_signedVC$, $A$ is the resolved public key from $Header.kid$, and $isValid$ is the boolean result of signature verification. The symbol $\perp$ denotes an invalid or failed verification result.



\begin{algorithm}
    \caption{VC Generation and Validation}\label{alg:vc_generation}
    \begin{algorithmic}[1]
    \Procedure{Generate\_VC}{$DID_{\text{issuer}}, \text{credential\_data}$}
        \State $VC \gets \{\text{id}: \text{baseURI} \parallel \text{uuidv4()},$
        \State \hspace{1.2cm} $\text{type}: [...],$
        \State \hspace{1.2cm} $\text{issuer}: DID_{\text{issuer}},$
        \State \hspace{1.2cm} $\text{credentialSubject}: \text{credential\_data} \}$
        \State $Header \gets \{\text{alg}: \text{"EdDSA"}, \text{typ}: \text{"JWT"}, \text{kid}: \text{key\_fragment} \}$
        \State $Payload \gets \text{Base64URLEncode}(VC)$
        \State $signingInput \gets \text{Base64URLEncode}(Header) \parallel \text{"."} \parallel Payload$
        \State $\sigma \gets \text{Ed25519Sign}(s_{\text{issuer}}, signingInput)$
        \State $J\_signedVC \gets signingInput \parallel \text{"."} \parallel \text{Base64URLEncode}(\sigma)$
        \State \Return $J\_signedVC$
    \EndProcedure
    
    \Procedure{Validate\_VC}{$J\_signedVC$}
        \State $(Header, Payload, \sigma) \gets \text{SplitJWT}(J\_signedVC)$
        \State $A \gets \text{ResolvePublicKey}(Header.kid)$
        \State $isValid \gets \text{Ed25519Verify}(A, Header \parallel \text{"."} \parallel Payload, \sigma)$
        \If{$isValid$}
            \State $VC \gets \text{Base64URLDecode}(Payload)$
            \State \Return $VC$
        \Else
            \State \Return $\perp$ \Comment{Invalid signature}
        \EndIf
    \EndProcedure
    \end{algorithmic}
\end{algorithm}

\subsubsection{\textbf{VP Generation and Validation}}
The process of generating and validating VP is described in Algorithm~\ref{alg:vp_generation}. $holder\_DID$ represents the decentralized identifier of the holder creating the VP. $credential\_JWTs$ denotes the set of signed verifiable credentials included within the VP. $nonce$ is a unique, verifier-provided random value used to prevent replay attacks, and $expiry$ specifies the VP’s validity duration. The $VP$ object contains fields such as an optional unique identifier ($\text{id}$), the holder’s DID ($\text{holder}$), the embedded credentials ($\text{verifiableCredential}$), the $nonce$, and the expiration timestamp ($\text{expiration}$). $Header$ defines the JWT header with signature algorithm (\texttt{EdDSA}), token type (\texttt{JWT}), and key identifier ($\text{kid}$). $Payload$ is the Base64URL-encoded $VP$, while $signingInput$ is the concatenation of the encoded header and payload. The signature $\sigma$ is generated by signing the $signingInput$ concatenated with $nonce$ using the holder’s private key $s_{holder}$. The signed JWT, $J\_signedVP$, combines the signing input and the encoded signature. In validation, $Header$, $Payload$, and $\sigma$ are extracted from $J\_signedVP$, and the decoded $VP$ is checked for expiration. Each embedded verifiable credential ($vc\_jwt$) is validated individually. The holder’s public key, $A_{holder}$, is resolved from the header key ID, and the signature is verified with $signingInput$ and the expected nonce. The symbol $\perp$ denotes an invalid or failed verification result.



\begin{algorithm}
    \caption{VP Generation and Validation}\label{alg:vp_generation}
    \begin{algorithmic}[1]
    \Procedure{Generate\_VP}{$holder\_DID$, $credential\_JWTs$, $nonce$, $expiry$}
        \State $VP \gets \{$
        \State \hspace{1.2cm} $\text{id}: \text{optional\_uuid}(),$
        \State \hspace{1.2cm} $\text{holder}: holder\_DID,$
        \State \hspace{1.2cm} $\text{verifiableCredential}: credential\_JWTs,$
        \State \hspace{1.2cm} $\text{nonce}: nonce,$
        \State \hspace{1.2cm} $\text{expiration}: \text{current\_time} + expiry$
        \State \} 
        \State $Header \gets \{\text{alg}: \text{"EdDSA"}, \text{typ}: \text{"JWT"}, \text{kid}: key\_fragment\}$
        \State $Payload \gets \text{Base64URLEncode}(VP)$
        \State $signingInput \gets \text{Base64URLEncode}(Header) \parallel "." \parallel Payload$
        \State $\sigma \gets \text{Ed25519Sign}(s_{holder}, signingInput \parallel nonce)$
        \State $J\_signedVP \gets signingInput \parallel "." \parallel \text{Base64URLEncode}(\sigma)$
        \State \Return $J\_signedVP$
    \EndProcedure

    \Procedure{Validate\_VP}{$J\_signedVP$, $expected\_nonce$}
        \State $(Header, Payload, \sigma) \gets \text{SplitJWT}(J\_signedVP)$
        \State $VP \gets \text{Base64URLDecode}(Payload)$
        \If{$\text{current\_time} \ge VP.\text{expiration}$}
            \State \Return $\perp$ \Comment{Expired presentation}
        \EndIf
        \ForAll{$vc\_jwt$ \textbf{in} $VP.\text{verifiableCredential}$}
            \If{$\text{Validate\_VC}(vc\_jwt) \neq \text{valid}$}
                \State \Return $\perp$
            \EndIf
        \EndFor
        \State $A_{holder} \gets \text{ResolvePublicKey}(Header.kid)$
        \State $isValid \gets \text{Ed25519Verify}(A_{holder}, signingInput \parallel expected\_nonce, \sigma)$
        \State \Return \textbf{if} $isValid$ \textbf{then} $VP$ \textbf{else} $\perp$
    \EndProcedure
    \end{algorithmic}
\end{algorithm}

\section{Performance Evaluation}
\label{sec:implementation-and-performance}
We developed several key components as a proof of concept, including the trust architecture, issuer, and source node, all implemented using JavaScript and Node.js. The trust architecture was designed to facilitate secure data validation, while the issuer and source node components were developed to handle the authentication and verification of data sources. To simulate a decentralized ledger, we initially set up the Tangle in a local environment, allowed us to test and refine how data was anchored and validated within a controlled environment. Once the data was successfully anchored in the Tangle, it was then moved to MongoDB for persistent storage. The transition from the Tangle to MongoDB was essential for ensuring long-term storage and retrieval of validated data. MongoDB was selected for its ability to manage large volumes of data with high performance. It is particularly well-suited for the trust layer, where data is continuously submitted by authenticated sources.  

The experiment was conducted to evaluate the performance of the proposed architecture under controlled conditions. The system was deployed on a local server powered by an Intel Core i7-1165G7 @2.80GHz (8 cores) processor, 16 GB of RAM, and running on Kali Linux. To ensure reproducibility, all configurations, input payloads, and system logs were carefully preserved throughout the evaluation. Performance metrics were collected using Autocannon for load testing and the built-in system monitor for hardware-level observation. Subsequent data analysis was performed using Python. The evaluation utilized a dataset provided by IBM, containing simulated financial transaction records. Each record includes detailed attributes such as the transaction timestamp, source and destination bank accounts, amount received, amount paid, the corresponding currencies, and the payment format.

\subsection{Experimental Results}
\label{sub:performance-evaluation}
The performance evaluation of the trust layer implementation is designed to examine how effectively the system upholds data trustworthiness without introducing significant delays in overall processing time and how effective the system manage resources. To achieve this, metrics are collected across three scopes: individual core layers (Layer 1, Layer 2, and Layer 3), the end-to-end transaction, and a baseline derived from existing open banking protocols. Each performance metric focuses on at least one of these scopes, depending on its relevance.

\begin{figure}[t]
    \centering
    \includegraphics[width=1\linewidth]{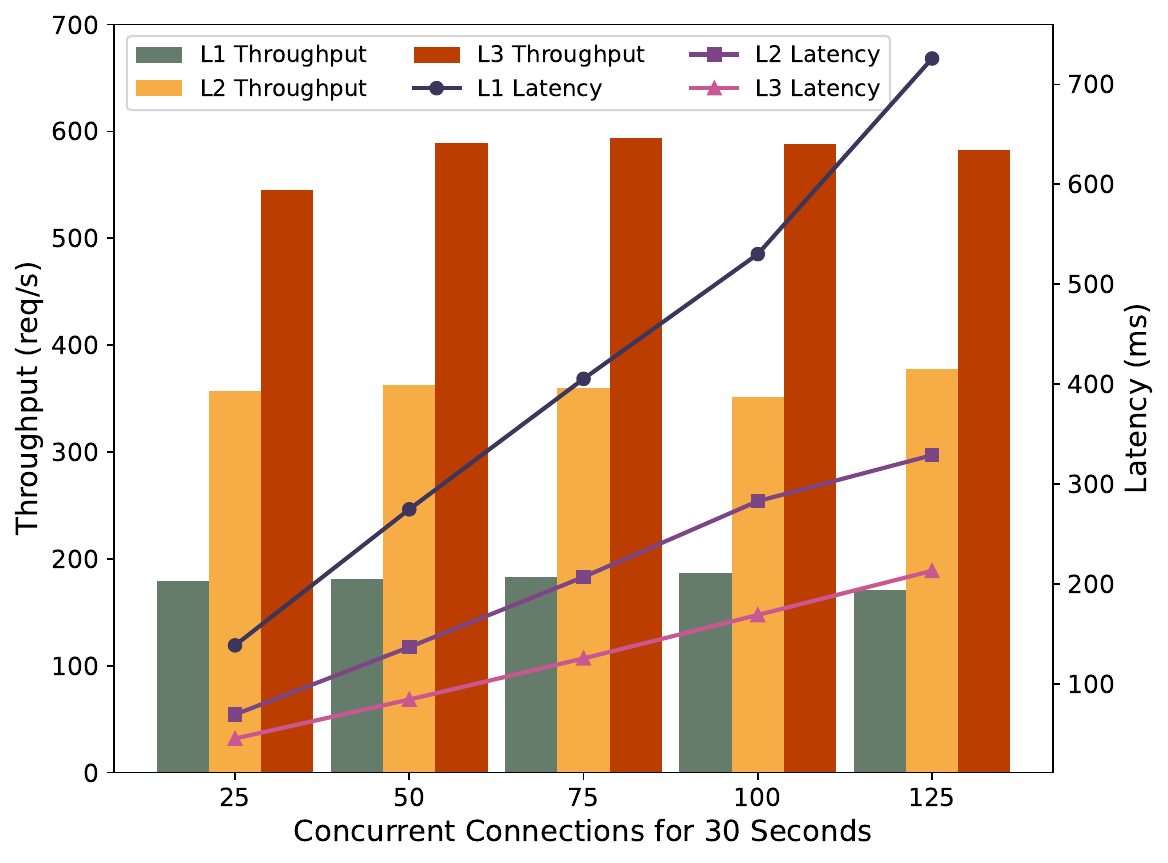}
    \caption{Throughput and latency for each layer, where each performs a distinct validation task, are measured under increasing concurrent connections. Showing how each layer contribute to the overall system performance.}
    \label{fig:avg-each-layers}
\end{figure}

\subsubsection{Throughput and Latency}
\label{subsub:througphut and latency}
\begin{figure}[t]
    \centering
    \includegraphics[width=1\linewidth]{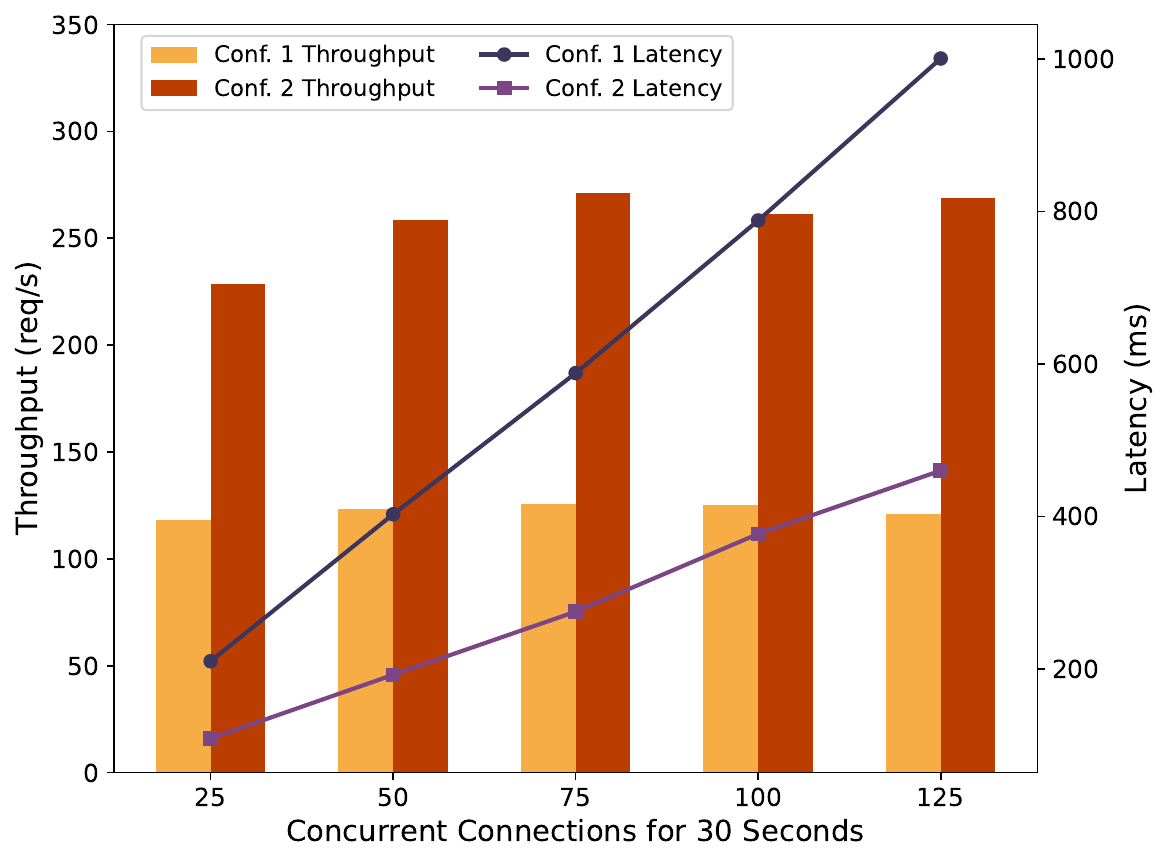}
    \caption{Throughput and latency for end-to-end interaction under two configurations: Configuration 1, where the data source is verified on every request, and Configuration 2, where the data source is pre-verified.}
    \label{fig:throughput-lat-end-to-end}
\end{figure}

Throughput and latency were evaluated from two perspectives: individually at each core layer (Layer 1, Layer 2, and Layer 3) and across the full end-to-end transaction path, as shown in Fig.~\ref{fig:avg-each-layers} and Fig.~\ref{fig:throughput-lat-end-to-end} respectively. The system was tested under varying levels of concurrency to assess its scalability and responsiveness. Specifically, we conducted tests with 25, 50, 75, 100, and 125 concurrent transactions over a fixed duration of 30 seconds. All of these scenarios reflect the intended design, which prioritizes the efficient handling of large, structured payloads per transaction over high-frequency, lightweight transactions.

Each layer analysis as shown in Fig.~\ref{fig:avg-each-layers} shows that throughput relatively constant at each layer for different number of concurrent connections, indicating the system’s ability to maintain steady processing rates as traffic increases. Latency trends, however, reveal distinct differences between the layers. Layer 2 and Layer 3 maintain relatively stable latency as concurrency increases, indicating efficient processing logic or lightweight operations. In contrast, Layer 1 exhibits a noticeable increase in latency, particularly beyond 75 concurrent connections. This suggests that Layer 1, likely responsible for more resource-intensive operations that becomes the performance bottleneck under high load.

\begin{figure*}[t]
    \centering
    \begin{tabularx}{\linewidth}{XXX}
        \includegraphics[width=0.32\textwidth]{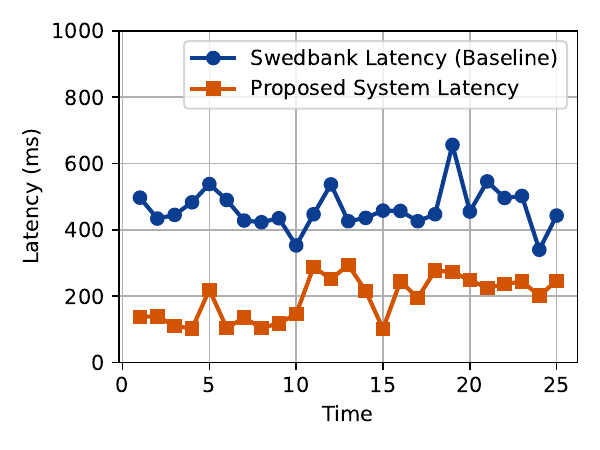}
        \caption{Latency comparison for the data query process between the proposed system and Swedbank. Proposed system achieves lower latency, despite performing additional hash validation.}
        \label{fig:latency-vs-baseline}
        &
        \includegraphics[width=0.32\textwidth]{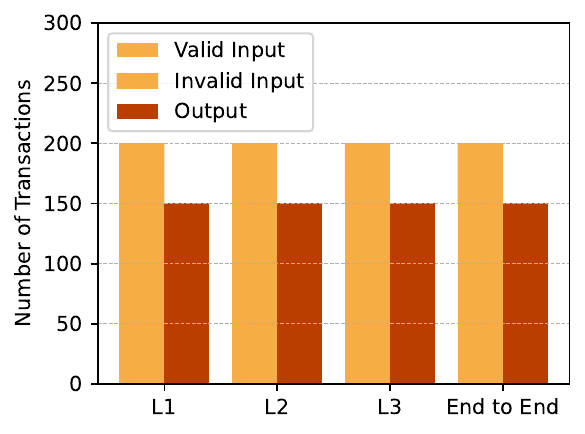}
        \caption{Comparison between valid and invalid input transactions and the resulting system output. Showing 100\% success rate to reject invalid input.}
        \label{fig:valid-invalid-output}
        &
        \includegraphics[width=0.32\textwidth]{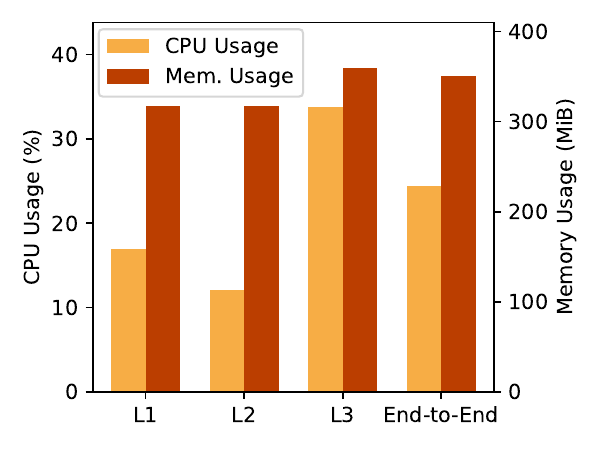}
        \caption{Comparison of CPU and memory usage across each layer and the end-to-end transaction. Showing end-to-end transaction is affected by each layer.}
        \label{fig:cpu-mem}
    \end{tabularx}
\end{figure*}

End to end analysis as shown in Fig.~\ref{fig:throughput-lat-end-to-end} highlights how differences in verification placement within the system architecture significantly affect scalability and responsiveness. In Configuration 1, data sources must first be verified through Layer 1 before data can proceed further in the pipeline. Meanwhile, Configuration 2 assumes that data sources have already been verified in Layer 1 and thus begins processing directly from Layer 2. Throughput analysis shows that Configuration 2 able to maintain high throughput as concurrent connections increase. Latency patterns further reinforce the efficiency of Configuration 2. While Configuration 1 starts with moderate latency (around 200 ms at 25 concurrent users), it escalates sharply beyond 900 ms at peak load (125 connections). In contrast, Configuration 2 maintains a more stable latency curve, remaining under 500 ms even under high concurrency. This suggests that the bottleneck introduced by real-time Layer 1 verification in Configuration 1 becomes increasingly detrimental as system load intensifies.

\subsubsection{Latency Compared to Baseline}
\label{subsub:latency-vs-baseline}
Fig.~\ref{fig:latency-vs-baseline} shows the latency comparison between the proposed trust layer and the baseline system (Swedbank). The comparison was tested on data query use case and the baseline data is gathered from Swedbank documentation~\cite{swedbank2025api}. The comparison reveals significant performance improvements over time. As shown in the figure, the proposed system consistently maintains lower latency throughout the entire 25-second observation window. The Swedbank baseline begins with a latency of approximately 500 ms and having peak at around 620 ms. In contrast, the proposed system starts under 200 ms and remains below 400 through the entire observation window.

\subsubsection{Input and Output Transaction}
\label{subsub:input-output}
Fig.~\ref{fig:valid-invalid-output} shows the error rate analysis across all layers and the end-to-end pipeline. The results demonstrates a flawless validation mechanism, with each layer achieving a 100\% success rate in processing only valid transactions. Every invalid input is effectively filtered and excluded, resulting in output transaction counts that precisely match the valid input numbers at each layer.

\subsubsection{CPU and Memory Usage}
\label{subsub:cpu-mem-usage}
Fig.~\ref{fig:cpu-mem} shows the resource utilization metrics, including CPU and memory usage. The utilization was recorded with 125 concurrent transactions for 30 seconds. The result indicates that CPU usage remains low throughout the stack, peaking under 35\% even in the most resource-intensive layer. Memory usage is equally well-controlled, staying within a tight band of 100–350 MiB across all layers. Notably, even in the full end-to-end scenario, memory consumption remains well below critical thresholds.

\section{Conclusion}
\label{sec:conclusion}
This paper presented the design and evaluation of layered trust architecture, built to ensure data integrity, low-latency access, and scalable ingestion in financial industry. Experimental results across core dimensions, including throughput, latency, success rate, and resource utilization, demonstrate the effectiveness of proposed design. Performance analysis shows that the latency scales linearly and throughput remain stable with increased concurrency. Trust enforcement mechanisms proved to be precise, with each layer maintaining a 100\% validation success rate. Furthermore, the system's lightweight CPU and memory footprint (under 35\% CPU and 350 MiB memory) highlights its deployability even in constrained environments. In conclusion, this paper offers a practical and efficient trust layer for modern open banking framework. Future work may optimize the resource hungry layer in our proposed trust layer, explore dynamic trust adaptation, and distributed deployment across edge devices.

\section*{Acknowledgment}
\noindent This work was supported by Hibah Penelitian Fundamental (PFR), Ministry of Higher Education, Science, and Technology Indonesia, grant number 048/E5/PG.02.00.PL/2024 - 2679/UN1/DITLIT/PT.01.03/2024.

\bibliography{references}{}
\bibliographystyle{IEEEtran}

\end{document}